\documentclass[preprint,prd,nofootinbib,superscriptaddress 12 pt]{revtex4}
\pdfoutput=1
\def\gsim{ \lower .75ex \hbox{$\sim$} \llap{\raise .27ex
\hbox{$>$}} }
\def\lsim{ \lower .75ex \hbox{$\sim$} \llap{\raise .27ex
\hbox{$<$}} }

\usepackage{graphicx}
\usepackage{dcolumn}
\usepackage{bm}
\usepackage{graphicx}
\usepackage{color}
\usepackage{amssymb}
\usepackage{bbm}
\usepackage{graphicx, amsmath, amssymb, setspace, color, slashed}

\newcommand\ee{\end{equation}}
\newcommand\be{\begin{equation}}
\newcommand\eea{\end{eqnarray}}
\newcommand\bea{\begin{eqnarray}}

\def\beq{\begin{equation}}
\def\eeq{\end{equation}}

\def\barr{\begin{array}}
\def\earr{\end{array}}

\def\fnote#1#2{\begingroup\def\thefootnote{#1}\footnote{#2}
     \addtocounter{footnote}{-1}\endgroup}

\begin{document}
\title{Small Steps and Giant Leaps in the Landscape}

\author{Adam~R.~Brown$^{1, \, 2}$  and Alex~Dahlen$^{2}$}
\affiliation{$^{1\,}$Princeton Center for Theoretical Science, Princeton, NJ 08544, USA \\
$^{2 \,}$Physics Department, Princeton University, Princeton, NJ 08544, USA \vspace*{.52 in}}

\fnote{}{emails: \tt{adambro@princeton.edu, adahlen@princeton.edu}}

\begin{abstract}
For landscapes of field theory vacua, we identify an effect that can greatly enhance the decay rates to wildly distant minima---so much so that such transitions may dominate over transitions to near neighbors.   We exhibit these `giant leaps' in both a toy two-field model and, in the thin-wall limit, amongst the four-dimensional vacua of 6D Einstein-Maxwell theory, and it is argued that they are generic to landscapes arising from flux compactifications.  We discuss the implications for the cosmological constant and the stability of stringy de Sitter.
\end{abstract}

\maketitle
\section{Introduction} \label{secI}

A field at $\phi_0$ is classically stable, but quantum mechanically unstable: it may tunnel out by nucleating a bubble of lower vacuum. But which lower vacuum? There's lots of them. Does it take a small step, or a giant leap?
 
 \begin{figure}[htbp] 
    \centering
    \includegraphics[width=5in]{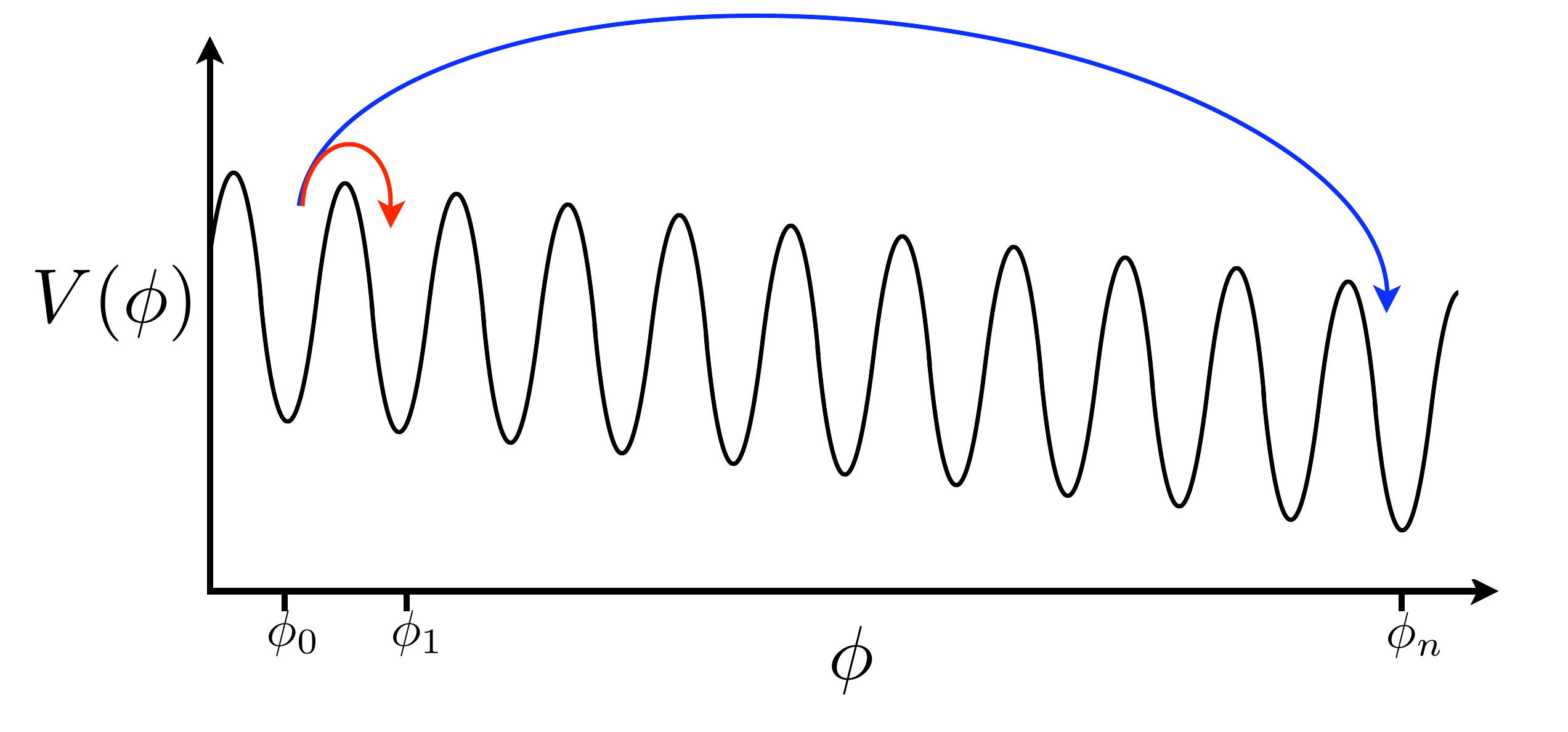} 
    \caption{\textcolor{red}{Small step} or \textcolor{blue}{giant leap}? For this potential, the answer is small step.}
    \label{fig1}
 \end{figure}
 
For the potential of Fig.~1---a single field with evenly spaced minima and identical barriers---the answer is a small step. In the thin-wall limit [equivalent, here, to tiny $\epsilon_n \equiv V(\phi_n) - V(\phi_0)$], the rate to tunnel directly to the $n^{\text{th}}$ vacuum, $\Gamma_n$, is  \cite{Coleman:1977py}
\begin{eqnarray}
\Gamma_n &\sim & e^{-B_n / \hbar}  \\
B_n & = & \frac{27 \pi^2}{2} \frac{\sigma_n^{\hspace{1pt} 4}}{\epsilon_n^{\hspace{2pt} 3}}, \label{eq:2}
\end{eqnarray}
where $\sigma_n$ is the tension of the domain wall separating the two phases,
\begin{equation}
\sigma_n \equiv \int_{\phi_0}^{\phi_n} d\phi \sqrt{2 [V(\phi)- V(\phi_n)]}. \label{eq:tension}
\end{equation}
It follows that, since both $\epsilon$ and $\sigma$ scale linearly with the distance jumped, 
\begin{eqnarray}
\epsilon_n &  = & n \, \epsilon_1, \textrm{ and}   \label{eq4}\\
\ \ \sigma_n & = & n \, \sigma_1,  \label{eq5}
\end{eqnarray}
then so too does the tunneling exponent
\begin{equation}
 B_n  =  n B_1. \label{eq:Bn} \ \ \ \ 
\end{equation}
For this potential, in this limit, you are exponentially most likely to tunnel to the first available vacuum.

But this isn't always so. In this paper we will add a second field direction, orthogonal to the line of minima, along which the potential asymptotes to zero. We will show that $\sigma_n$ may then grow slower than $n^{3/4}$ so that the most rapid transitions are to wildly distant minima.  

 \begin{figure}[htbp] 
    \centering
    \includegraphics[width=5in]{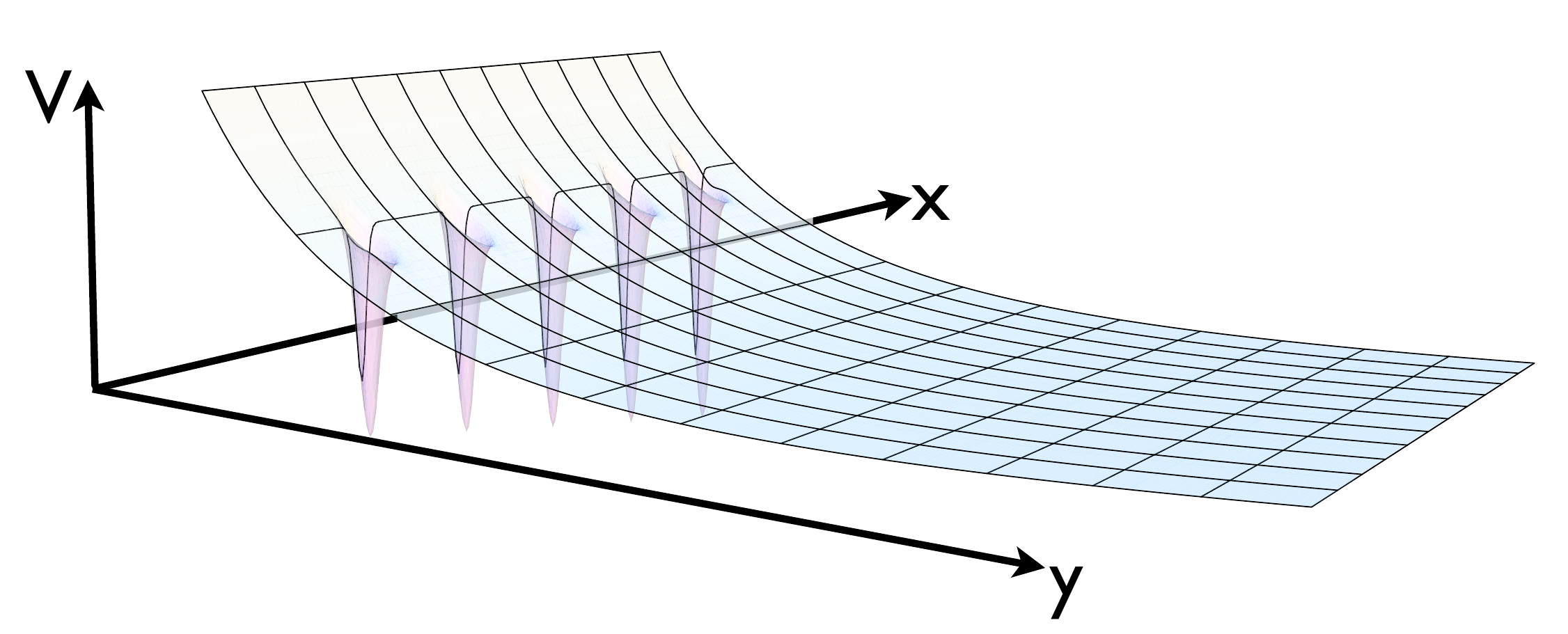} 
    \caption{A two-field potential with a ladder of ever-deeper minima and a second, orthogonal field direction in which the field asymptotes to zero. Along the line of minima, the potential resembles Fig.~\ref{fig1}. In Sec.~\ref{secIII} we will see that the full potential resembles the effective potential of a flux compactification.}
    \label{fig2}
 \end{figure}

In Sec.~\ref{secII} we will give a precise characterization of these giant leaps\footnote{While this effect shares a common conclusion with \cite{Tye:2009rb} (resonance tunneling), \cite{Brown:2007zzh} (antibrane assisted tunneling), and \cite{Feng:2000if} (exponential prefactors)---that some transitions in the landscape are much faster than you might naively imagine---the effect itself is quite distinct.  For giant leaps in a different setting, see \cite{Dienes:2008qi}.} in the context of a simple two-field model. In Sec.~\ref{secIII} we will find giant leaps amongst the four-dimensional vacua of 6D Einstein-Maxwell theory, the simplest flux compactification with de Sitter minima \cite{BlancoPillado:2009di, Freund:1980xh}; the size of the extra dimensions will function as the crucial second field direction. (We will also be aided by the fact that, for this theory, $\epsilon_n$ grows faster than linearly with $n$.)  In Sec.~\ref{secIV} we will argue for the existence of giant leaps in general flux compactifications, and discuss their relevance for the cosmological constant problem. By opening up new decay channels, and fast ones, giant leaps pose a threat to the stability of stringy de Sitter.

\section{A toy model} \label{secII}
In this section, we are going to study tunneling in a two-field theory designed to be the simplest possible model rich enough to manifest the phenomenon of interest.

Consider two 3+1-dimensional fields, $x$ and $y$, and a potential that is given by
\begin{equation}
V(x,y) = \frac{1}{y^2} \label{eq:pot}
\end{equation}
save for a ladder of evenly-spaced, ever-deeper, essentially-pointlike minima along the line $y=1$,
\begin{eqnarray}
V(x=0, y=1)  & = & E,  \ \ \, \ \ \ \ \ \ \ \ \, \, \, \hspace{1pt} \ \textrm{ the false vacuum, and} \nonumber \\
V(x = n, y=1) & = & E - \epsilon_1 \, n \ \, \ \ \ \ \   \ \textrm{for $n \in \mathbb{N}$, the truer vacua}.
\end{eqnarray}
This potential is plotted in Fig.~\ref{fig2}. Were the field restricted to the line $y=1$, this potential would look a lot like the potential of Sec.~\ref{secI}: $\epsilon_n = n \epsilon_1$ and (in the thin-wall limit) $\sigma_n = n \sigma_1$, so that $B_n = n B_1$. But the field is not restricted to $y=1$, and that will make all the difference.

To have a minimum at all, $E$ must be less than $1$. For $E\le0$, the only option is to move down the ladder; for $0<E<1$, tunneling out to $y = \infty$ is also possible. We will study both signs of $E$ eventually, but first we will consider the case for which the answer is most simple, the effect most vivid, and the leaps most gigantic: thin-wall decays from $E=0$. 
  
\subsection{The basic idea}

Decay is mediated by an O(4)-symmetric instanton that describes the formation of a bubble of pure true vacuum separated from the ambient pure false vacuum by an interpolating wall. For tiny enough $\epsilon_n$, the wall is thin and its profile is given by the path that minimizes Eq.~\ref{eq:tension}. This optimal path is determined by a tradeoff: straight paths that follow $y=1$ are shorter, while loopy paths that detour to large $y$ are longer, but see a lower potential. 

\begin{figure}[htbp] 
    \centering
    \includegraphics[width=3.9in]{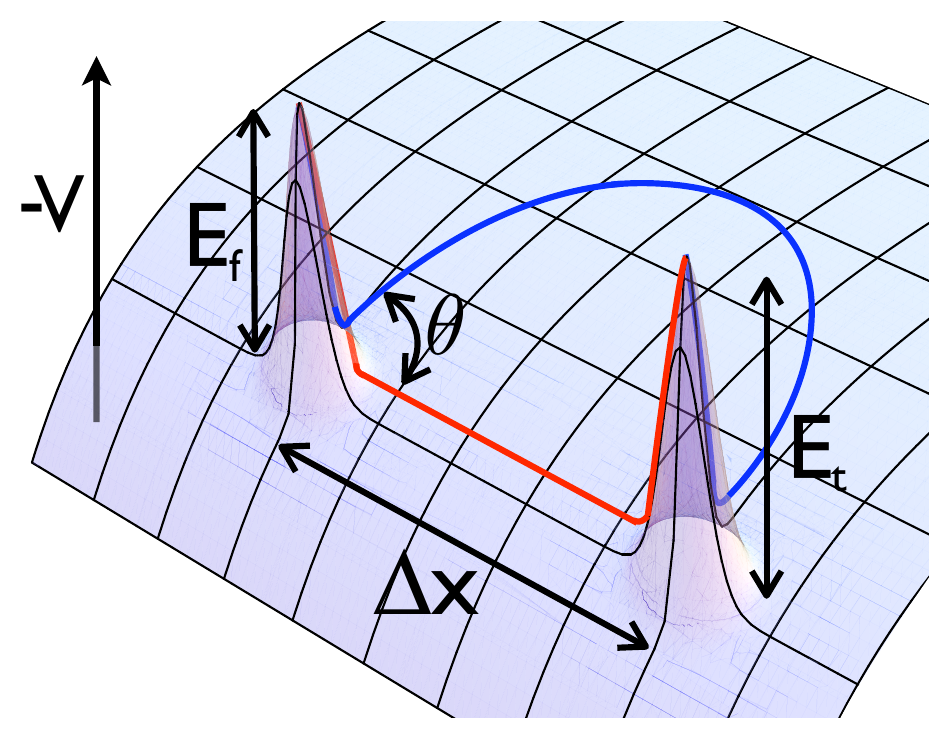} 
    \caption{What goes up must come down. In interpolating between the two vacua, the field in the wall does not follow a \textcolor{red}{straight line} in field space, instead it \textcolor{blue}{loops} up to large $y$ and then back down to $y=1$.}
    \label{fig3}
 \end{figure}

Minimizing the wall tension is mathematically equivalent to solving for the trajectory of a frictionless ball moving in a potential $-V(x,y)$. The ball leaves the false vacuum with a velocity set by the energy, $\frac{1}{2} v^2 = 1 - E$, and must be aimed so as to land precisely on the true vacuum, as in Fig.~\ref{fig3}. In this way of looking at things, the ball is aimed `upwards' and then curves back down under the influence of the slope.

Were the potential linear, the ball's trajectory would be a parabola. Instead, for this potential, when $E = 0$, the ball follows an arc of a circle whose diameter lies along $y=0$ and which connects the vacua at $(x=0,y=1)$ and $(x=n,y=1)$, 
\begin{equation}
(x + \textrm{const.})^2 + y^2 = \frac{1}{\cos^2 \theta}, \label{eq:circle}
\end{equation}
where $\theta$ is the angle to the horizontal at $y=1$. 

Between leaving $y=1$ on the way up and arriving again at $y=1$ on the way down, the field traverses 
\begin{equation}
\Delta x   =  2 \tan \theta.
\end{equation}
For $E=0$, the ball has just enough velocity to escape, and can travel arbitrarily far by starting with $\theta$ arbitrarily close to $\pi/2$.  (Strictly speaking, we are only interested when $\Delta x = n \in \mathbb{N}$, so that the ball lands in an awaiting vacuum.)

We calculate the tension of the domain wall by inserting its profile, Eq.~\ref{eq:circle}, into the WKB formula, Eq.~\ref{eq:tension}:
\begin{equation}
\sigma_{n}  =  2 \sqrt2 \, \textrm{arcsinh} \frac{n }{2} \label{eq:sigma}.
\end{equation}
Though the tension grows monotonically with $n$, it does so much slower than linearly. This is easy to understand. Getting from the false vacuum to the true vacuum can be roughly divided into three phases: first tunnel out to large $y$, second tunnel along in $x$, third tunnel back to $y=1$. But by the time you have gone to all the effort/incurred all the action of tunneling out in $y$, tunneling a little further in $ x $ doesn't really cost you much more. 

 \begin{figure}[h] 
    \centering
    \includegraphics[width=5in]{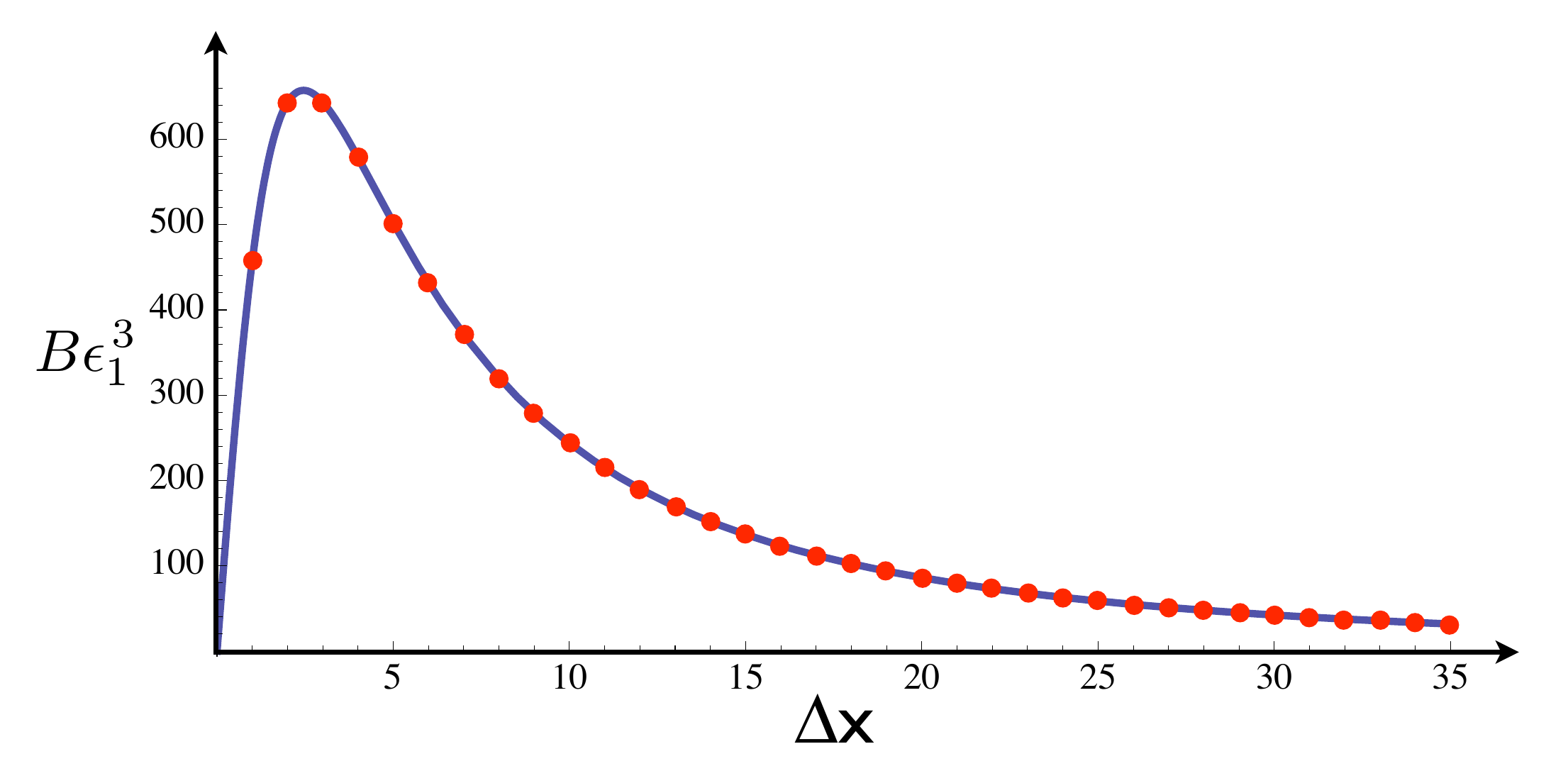} 
    \caption{The tunneling exponent (scaled by $\epsilon_1^{\hspace{2pt} 3}$), in the thin-wall limit, starting from the $E=0$ vacuum. The likeliest tunneling event is a giant leap far down the landscape.  Only integer values of $\Delta x$, indicated by red dots, count. (In subsequent figures, we will omit the dots.)}
    \label{fig4}
 \end{figure}
 
As a consequence the tunneling exponent of Eq.~\ref{eq:2}, plotted in Fig.~\ref{fig4}, is now smallest for the most distant minima.  For this potential, in this limit, starting from the $E=0$ vacuum, you are exponentially most likely to tunnel to the farthest available vacuum.
 
 \subsection{Away from $E = 0$}
For $E \neq 0$, but still in the thin-wall limit, the circle of Eq.~\ref{eq:circle} stretches to an ellipse,
\begin{equation}
\frac{1 - \frac{1}{2} v^2 \sin^2 \theta}{\frac{1}{2} v^2 \cos^2 \theta} (x + \textrm{const.})^2 +  y^2  =  \frac{1}{1 - \frac{1}{2} v^2 \sin^2 \theta} \label{conic},
\end{equation}
whose (minor for $E>0$, major for $E<0$) axis still lies along $y=0$ and which connects the vacua at $(x=0,y=1)$ and $(x=n,y=1)$. In this case,
\begin{equation}
\Delta x = \frac{\frac{1}{2} v^2 \sin 2 \theta }{1 - \frac{1}{2} v^2 \sin^2 \theta}.
\end{equation}
(For small $v$, the potential is approximately linear, and we recover the familiar result, $\Delta x = v^2 \sin 2 \theta /g$, for a ball subject to gravity $g=-V'(y=1) = 2$.) Again, we calculate the tension by inserting the domain wall profile, Eq.~\ref{conic}, into Eq.~\ref{eq:tension}:
\begin{equation}
\sigma  =  2\sqrt{2} \left[ \textrm{arcsinh}[\tan \theta] - (1-\frac{1}{2} v^2) \frac{ \frac{1}{\sqrt{2}} v \sin \theta}{1 - \frac{1}{2} v^2 \sin^2 \theta} \right] , \label{fulltension}
\end{equation}
which reduces to Eq.~\ref{eq:sigma} when $E = 1 - \frac{1}{2}v^2 = 0$.  

 \begin{figure}[htbp] 
    \centering
    \includegraphics[width=5in]{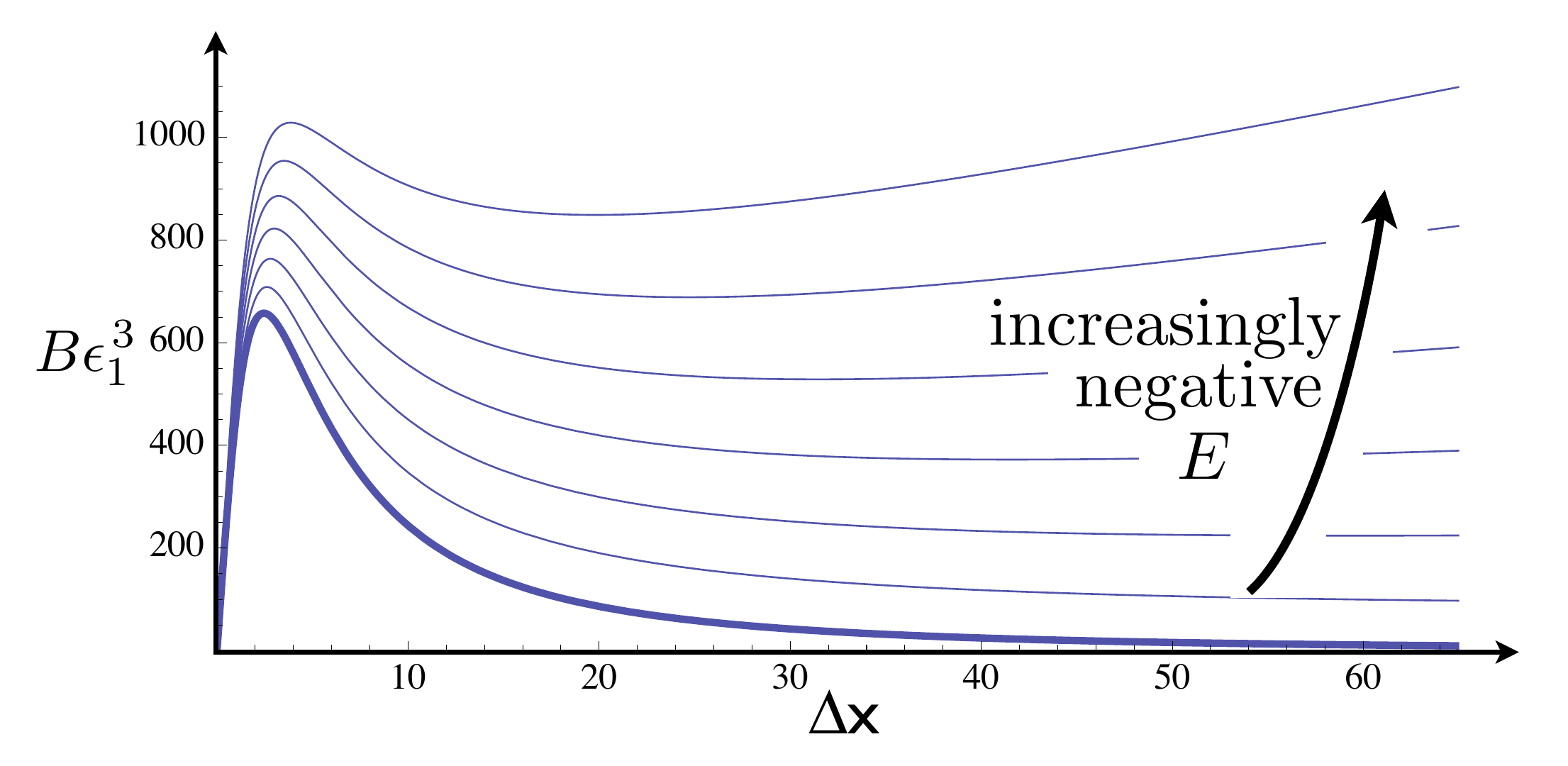} 
    \caption{The tunneling exponent, out of a vacuum with $E \leq 0$, in the thin-wall limit, for various values of $E$.  Shown are $E = 0$ (in bold), $-0.02, -0.04, \dotsc, -0.12$.  The bold line is the same as the line in Fig.~\ref{fig4}, and you can see how the solutions match on as $E\rightarrow0$ from below.  At large $\Delta x$, the curves no longer go to zero, but turn around and grow linearly.}
    \label{fig6}
 \end{figure}

{\bf For} $\mathbf{E<0}$, the tunneling exponent $B$ is plotted in Fig.~\ref{fig6}.  Unlike for $E=0$, $B$ reaches a minimum and then resumes growing with distance. This difference is because  the barrier height now never vanishes; no matter how far out you go in $y$, the smallest it gets is $-E$.  Negative $E$ also means that the ball's velocity $v$ exceeds the escape velocity for this potential. There is a critical value of $\theta$ past which the ball escapes, and its trajectory changes from an ellipse to a hyperbola; this does not correspond to a tunneling solution.  As $\theta$ approaches the critical value from below, $\Delta x$ can be made arbitrarily large.  While tunneling past the minimum of $B$ is highly suppressed, it is in principle possible to tunnel arbitrarily far. 

\begin{figure}[htbp] 
    \centering
    \includegraphics[width=\textwidth]{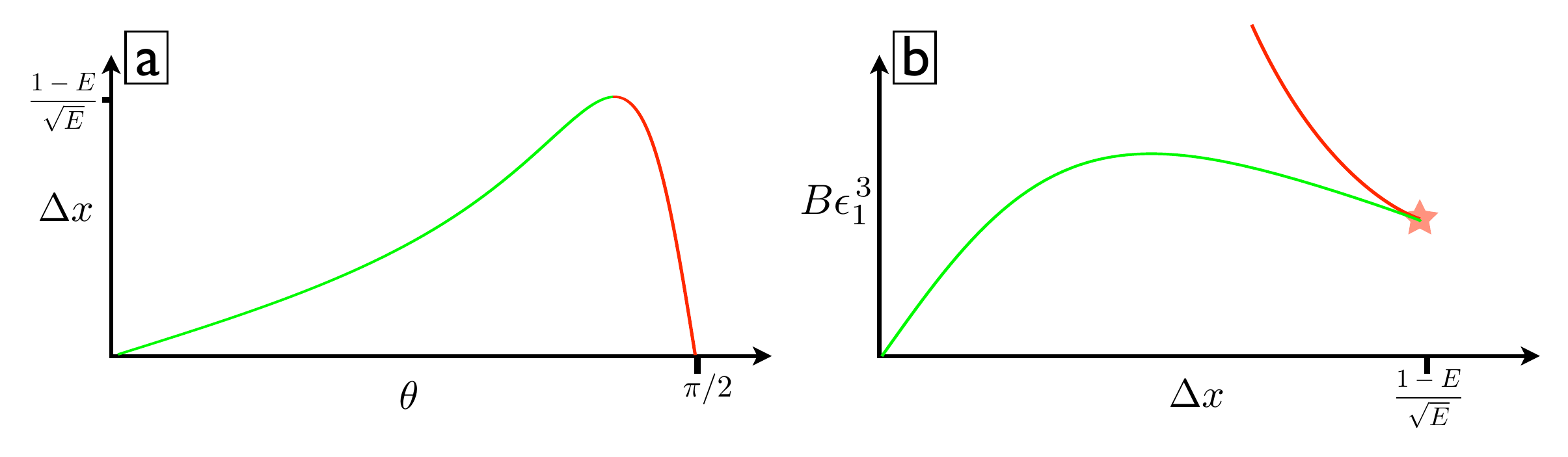} 
    \caption{(a) For $E>0$, the ball doesn't have escape velocity and can travel no farther than $(1-E)/\sqrt{E}$. Increasing $\theta$ past the optimal value makes the ball go higher but less far.  (b) Two instantons, a low road and a high road, contribute to the decay rate.  At $\Delta x_\text{max}$, they meet and annihilate.}
    \label{fig19}
 \end{figure}

{\bf For} $\mathbf{E > 0}$, this is not true.  The ball has less than the escape velocity, and there is a maximum distance it can travel, which corresponds to a maximum distance the field can tunnel.  Equation~\ref{conic} implies that for our potential this distance is
\begin{equation}
\Delta x _\text{max} = \frac{\frac{1}{2} v^2}{ \sqrt{1 - \frac{1}{2} v^2} } = \frac{1- E}{\sqrt{E}}.
\end{equation}
As $\theta$ is increased, $\Delta x$ increases for a while, but then hits a maximum; increasing $\theta$ further makes the ball go higher, but less far (see Fig.~\ref{fig19}a).  This is familiar from everyday life: if you throw a ball with fixed speed, there is a maximum distance you can throw it.

The tunneling exponent is plotted in Fig.~\ref{fig19}b.  For $\Delta x<\Delta x_\text{max}$, there are two instantons: a dominant one with small $\theta$ (green) and a subdominant one with large $\theta$ (red). In all subsequent plots we will only show the dominant branch.  For $\Delta x>\Delta x_\text{max}$, there are no instantons.  

The disappearance of these instantons is related to the existence of a second type of tunneling available to vacua with $E>0$. Rather than tunneling down the ladder in the $x$-direction, we can tunnel out in the $y$-direction. (In the 6D Einstein-Maxwell theory of the next section this will correspond to decompactification.)  Tunneling out is so dominant that it swallows instantons with $\Delta x>\Delta x_\text{max}$.  Any field configuration that tries to interpolate between two such distant vacua  is unstable---the wall will inevitably slide down the potential and roll out towards $y=\infty$ \cite{Aguirre:2009tp, Johnson:2008vn, Cvetic:1994ya}.  

\begin{figure}[htbp] 
    \centering
    \includegraphics[width=5in]{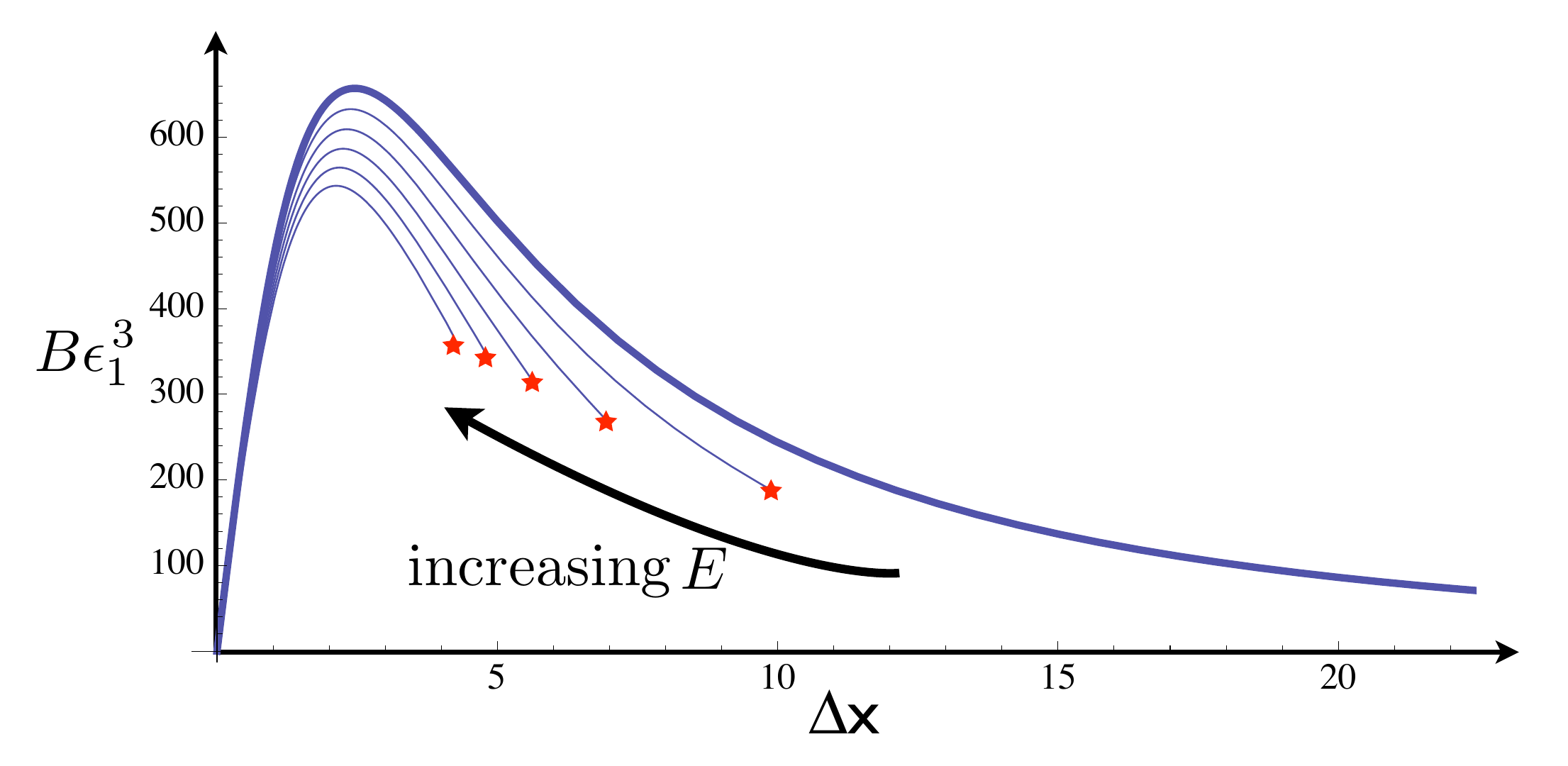} 
    \caption{The tunneling exponent, out of a vacuum with $E \geq 0$, in the thin-wall limit, for various values of $E$.  Shown are $E = 0$ (in bold), $0.01, 0.02, \dotsc, 0.05$.  For all values of $E>0$, the instanton disappears at some $\Delta x_\text{max}$, indicated by a small red star.  As $E\rightarrow0$, $\Delta x_\text{max}$ moves off to infinity.}
    \label{fig5}
 \end{figure}
The tunneling exponent $B$ is plotted in Fig.~\ref{fig5}.  The exponent rises, falls, and then abruptly stops. For larger $E$, `decompactification' is easier and $\Delta x_\text{max}$ is smaller. 

\subsection{Away from the thin-wall limit}

As $\epsilon_n$ gets bigger, the thin-wall approximation gets worse. This means that while our results become exact as $\epsilon_1 \rightarrow 0$, for any nonzero $\epsilon_1$ we cannot trust our results at large $n$. To truly understand giant leaps, we must abandon the thin-wall approximation. 

Though the transition is still mediated by an O(4)-symmetric instanton, the bubble wall is now thick. The rate exponent can no longer be approximated by Eq.~\ref{eq:2}, and we will instead need the full expression,
\begin{equation}
B = S_E(\textrm{instanton}) - S_E(\textrm{false vacuum}) = 2 \pi^2 \int d\rho \,\rho^3 \left\{ \frac{1}{2} \dot{x}^2 + \frac{1}{2} \dot{y}^2 + V(x,y) - E \right\}, \label{eq:B}
\end{equation}
the difference in Euclidean action between the instanton and the false vacuum. We can calculate this rate numerically. 

\begin{figure}[htbp] 
    \centering
    \includegraphics[width=5in]{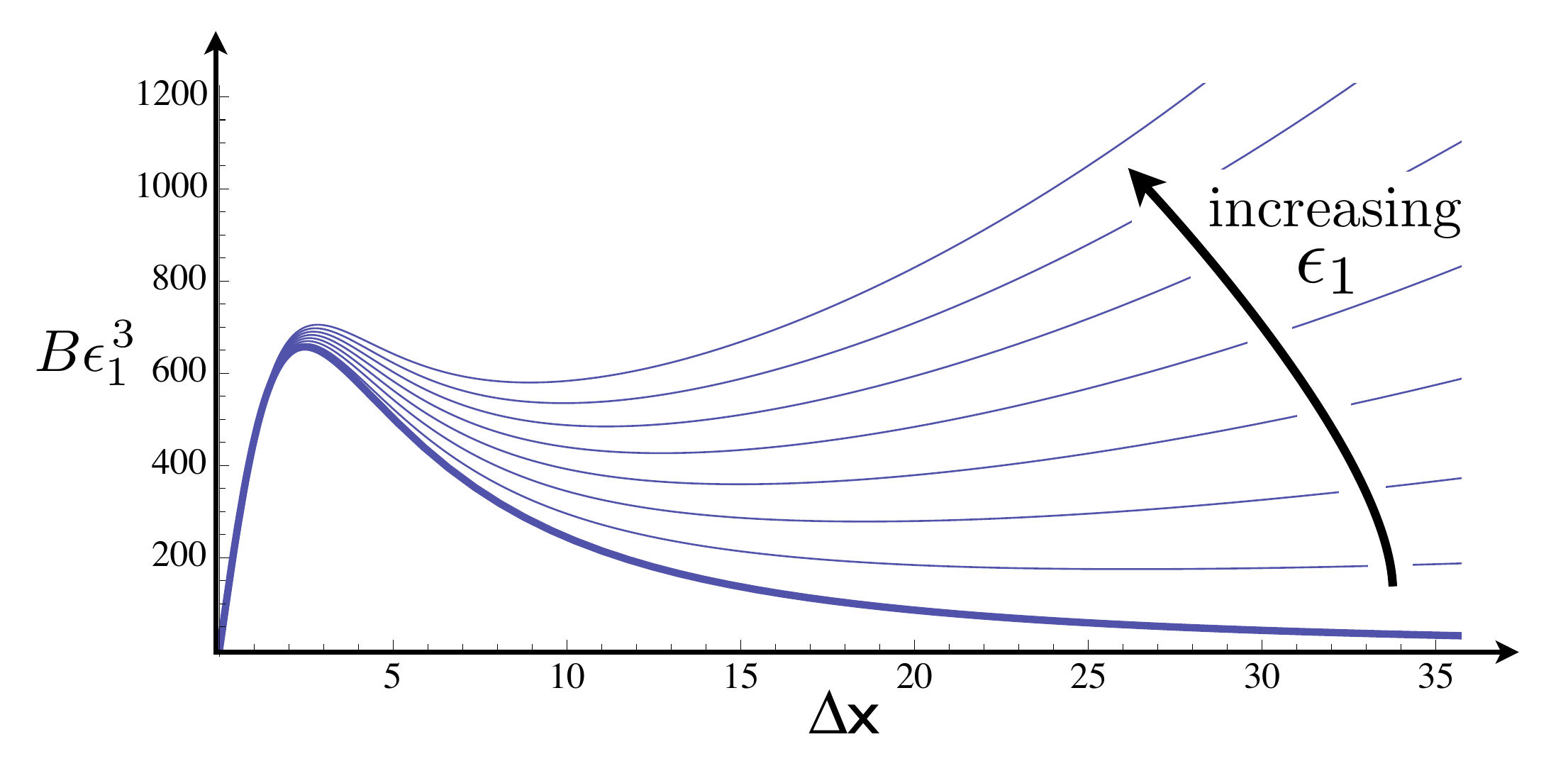} 
    \caption{Thick-wall corrections for $E=0$. The thin-wall limit corresponds to $\epsilon_1 = 0$, and is plotted in bold.  We also plot $\epsilon_1 = 0.002, 0.004, \dotsc, 0.014$.  Only in the thin-wall limit does $B$ fall forever; for any $\epsilon_1 > 0$, $B$ turns around and grows again. At larger values of $\epsilon_1$, deviations set in at smaller $\Delta x$. }
    \label{fig7}
 \end{figure}

{\bf For} $\mathbf{E=0}$ the result is shown in Fig.~\ref{fig7}. 
The thin-wall approximation breaks down for large $\Delta x$, and as we increase $\epsilon_1$, it breaks down ever sooner.  In the thin-wall approximation, we found that tunneling infinitely far is completely unimpeded.  However, for any nonzero $\epsilon_1$, no matter how small, the plot of $B$ now curves up, creating a minimum at a finite value of $\Delta x$.  As we increase $\epsilon_1$, the upswing happens earlier. For thick enough walls, the upswing happens so early that $B$ never even gets a chance to turn over: $B$ grows monotonically and giant leaps are exponentially suppressed.

\begin{figure}[h!] 
    \centering
    \includegraphics[width=5in]{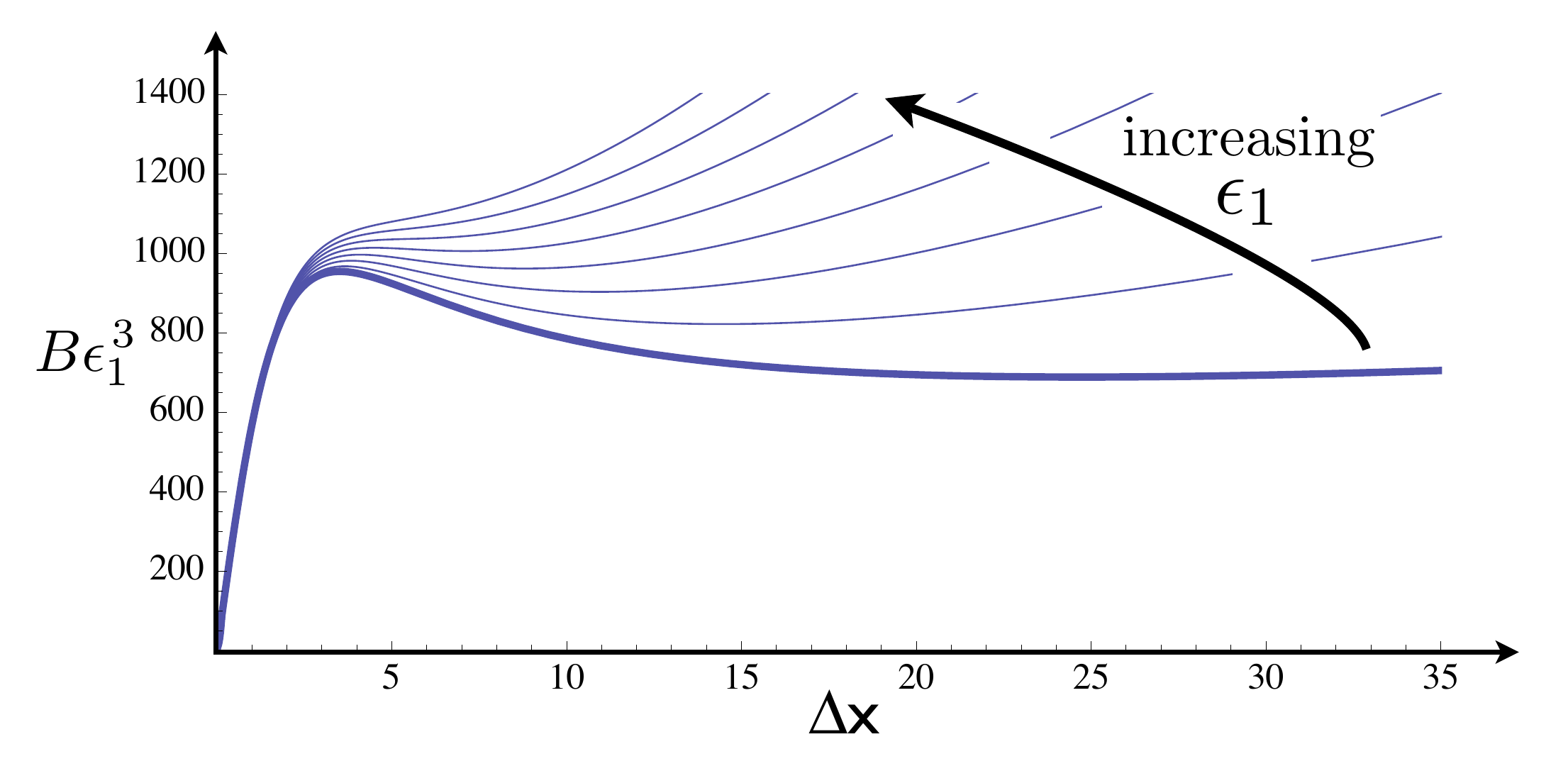} 
    \caption{Thick-wall corrections for $E<0$. Here, we have set $E=-0.05$, and have plotted $\epsilon_1 = 0$ (the thin-wall case, in bold), $0.002, 0.004, \dotsc, 0.014$.  Again, at larger values of $\epsilon_1$, deviations set in at smaller $\Delta x$. }
    \label{fig11}
 \end{figure}

{\bf For} $\mathbf{E<0}$ the result is shown in Fig.~\ref{fig11}, and there are no surprises. We have already seen that either making $E$ negative or making the wall thick causes the plot to curve up.  Here we see that doing both at once causes the plot to curve up all the more. 

\begin{figure}[htbp] 
    \centering
    \includegraphics[width=5in]{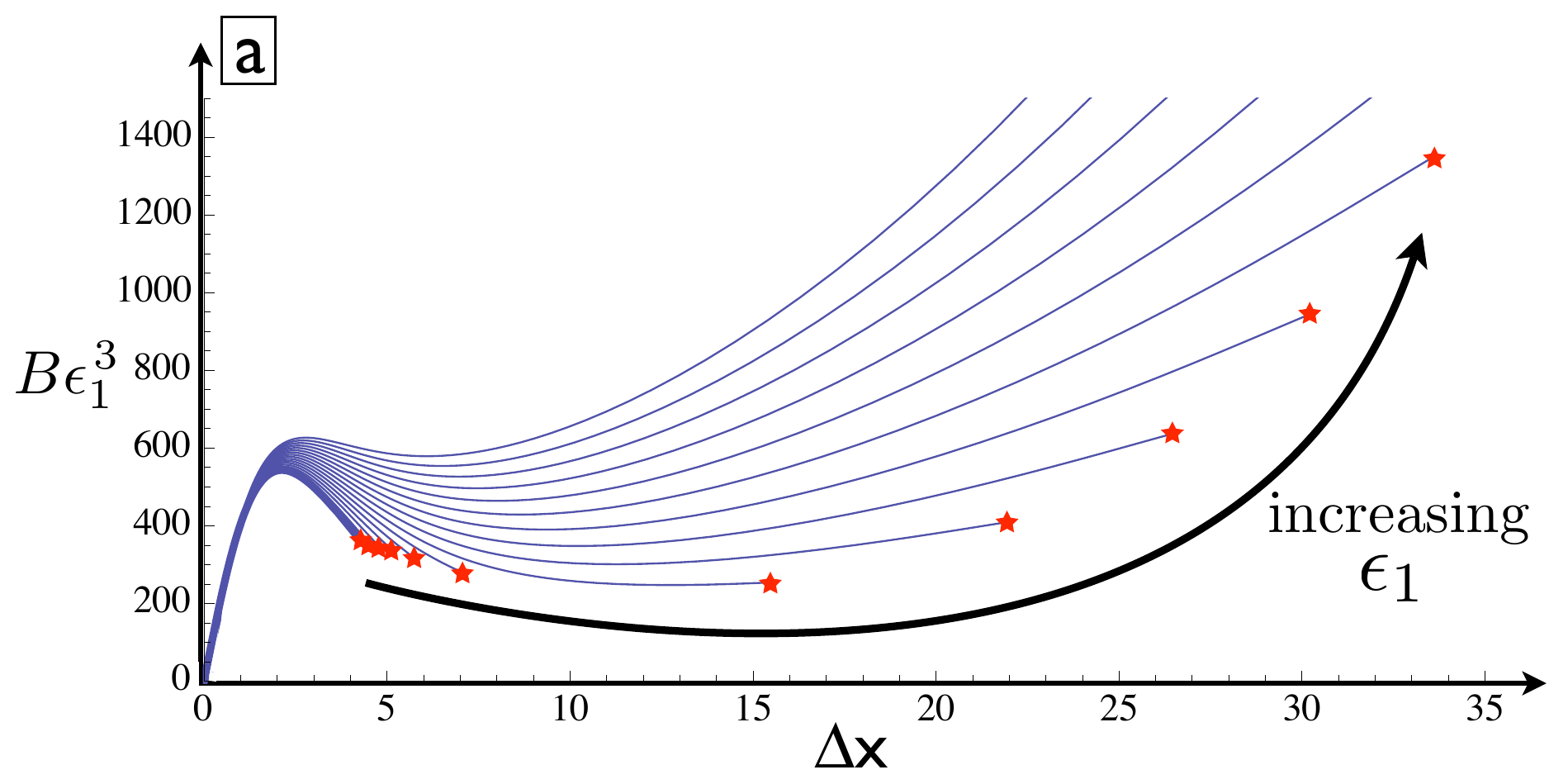} 
        \includegraphics[width=5in]{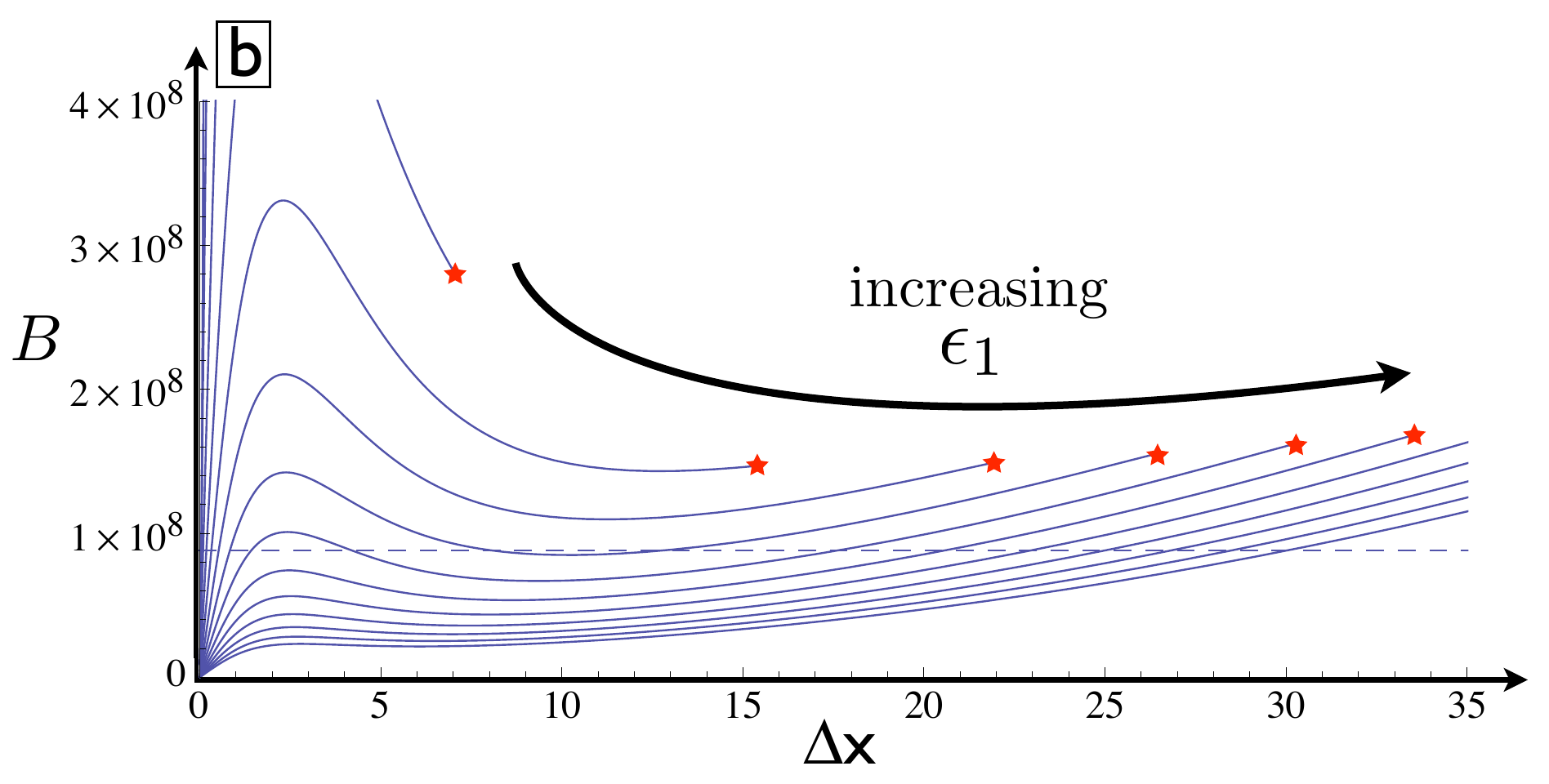} 
    \caption{Thick-wall corrections for $E>0$.  Here, we set $E=0.05$ and $\epsilon_1 = 0$ (the thin-wall case, in bold), $0.002, 0.004, \dotsc, 0.030$. In (a) we plot the scaled tunneling exponent $B\epsilon_1^{\,\,3}$.  Thickening the wall greatly increases $\Delta x_\text{max}$.  In (b) we plot the \emph{unscaled} tunneling exponent, $B$.  Removing the scaling inverts the order of the plots, $\epsilon_1=0$ is on top at $B=\infty$, and increasing $\epsilon_1$ lowers $B$.  The stars indicating $\Delta x_\text{max}$ line up in the two plots.  The decompactification transition is indicated by the dotted line, and there are cases where giant leaps beat both small steps and decompactification.  Note also that instantons always disappear with actions greater than the action to decompactify.}
    \label{fig8}
 \end{figure}

{\bf For} $\mathbf{E>0}$ the result is shown in Fig.~\ref{fig8}. Figure \ref{fig8}a shows the tunneling exponent scaled, as in previous figures, by $\epsilon_1^{\,\,3}$. We can see that increasing $\epsilon_1$ has two competing effects. As before, it curves the plot up for large $\Delta x$, discouraging giant leaps. But it also increases $\Delta x_{\text{max}}$, exposing new vacua and potentially lengthening the optimal jump. 

Figure \ref{fig8}b shows the same curves, except now unscaled. Removing the scaling has inverted the order of the plots from Fig.~\ref{fig8}a. Small $\epsilon_1$, which corresponds to the thin-wall limit, also corresponds to large $B$. We can now compare these rates against the rate to decompactify towards $y=\infty$, indicated by the dotted line of Fig.~\ref{fig8}b. There are two interesting things here.  First, and most important, there are cases where giant leaps not only beat small steps, but also beat decompactification.  Second, $B(\Delta x_{\text{max}})$ is always bigger than $B(\text{decompactification})$, for every value of $\epsilon_1$.  We believe this to be a general principle: only subdominant instantons may abruptly disappear\footnote{For tunneling from a pointlike $E=0$ vacuum in a background exponential potential, $V=e^{-y}$, it is amusing to follow the intricate way in which this principle is upheld. We would have a violation if there were a finite-action maximum tunneling distance $\Delta x_{\text{max}}$ (so that an instanton abruptly disappears) in a set-up in which decompactification is impossible, or at any rate higher in action (so that it is a \emph{dominant} instanton that abruptly disappears). Since the false vacuum is degenerate with the vacuum at infinity, decompactification \emph{is} impossible for field theory. For thick-walled field-theory instantons, the principle is respected because there is no  $\Delta x_{\text{max}}$. For thin-walled field-theory instantons, there \emph{is} a  $\Delta x_{\text{max}}$, but the principle is still upheld, because to get a bounded tunneling distance you need to go all the way to the thin-wall limit, which means sending $\epsilon_1$ to zero, which simultaneously sends the tunneling action ($B \sim \sigma^{n+1} / \epsilon^n$ in $n+1$ dimensions) to infinity. For quantum-mechanical instantons (i.e. $n=0$) there is a  $\Delta x_{\text{max}}$, and getting there requires only finite-action; but the principle is still upheld because, with a finite number of operative degrees of freedom, tunneling is now possible (and indeed dominant) to the degenerate decompactified vacuum. It is because of these subtleties, which are not present in the 6D Einstein-Maxwell theory, that we have expounded instead on the inverse quadratic potential.}.  We will discuss this further in an upcoming work. 

\subsection{Summary of the Toy Model}

A second field direction, orthogonal to the line of minima, along which the potential asymptotes to zero, allows $\sigma_n$ to grow slower than $n^{3/4}$ so that the most rapid transitions may be to wildly distant minima. The effect is strongest for small $\epsilon_1$ and near $E=0$, where giant leaps can beat both small steps and decompactification.

\section{6D Einstein-Maxwell Theory} \label{secIII}

Our next objective will be to show that giant leaps arise in landscapes resulting from flux compactifications.  We will start by considering in detail the tunneling between the four-dimensional vacua of 6D Einstein-Maxwell theory, as it is a simple model with many features relevant to the string landscape \cite{Douglas:2006es} and the mechanism of tunneling was recently worked out by Blanco-Pillado, Schwartz-Perlov, and Vilenkin (BSV) \cite{BlancoPillado:2009di}. 

\subsection{The model and its tunneling behavior}
We now give a brief review of the landscape of four-dimensional vacua of 6D Einstein-Maxwell theory, largely following BSV \cite{BlancoPillado:2009di}.

The action is
\begin{equation}
S_\text{EM}= \int d^{\,6}\!\,x \, \sqrt{-G}\left(\frac12 R^{(6)}-\frac14 F_{AB}F^{AB}-\Lambda_6\right), \label{eq:EMaction}
\end{equation}
where $A$ and $B$ run from $0$ to $5$, $R^{(6)}$ is the Ricci scalar associated with the metric $G_{AB}$, $F_{AB}$ is the Maxwell field strength, $\Lambda_6$ is a positive six-dimensional cosmological constant, and we use units for which $\hbar$, $c$, and the reduced 6D Planck mass are all 1.

Consider the sector of the theory in which two of the dimensions are compactified on a sphere and the metric takes the form
\begin{equation}
\label{metric}
ds^2=G_{AB}dx^A dx^B=\tilde{g}_{\mu\nu}dx^\mu dx^\nu + R^2 d\Omega_2^{\,\,2}.
\end{equation}
Here, $\tilde{g}_{\mu\nu}$ is the metric of four-dimensional de Sitter, Minkowski, or anti-de Sitter, and $d\Omega_2^{\,\,2}$ is the line element of a unit two sphere.  (The indices $\mu$ and $\nu$ run from $0$ to $3$ and cover the extended directions.)

The only ansatz for the Maxwell field that is consistent with the symmetries of the metric is the monopole configuration
\begin{equation}
F_{ab}=\frac{g N}{4\pi}\sqrt{g_2}\epsilon_{ab},
\end{equation}
where $g$ is the magnetic monopole charge, $\sqrt{g_2}\epsilon_{ab}$ is the volume form on a unit two-sphere, and all other components of $F_{AB}$ are zero.  We will follow \cite{Freund:1980xh} and assume there is a fundamental electric charge $e$, which imposes a Dirac quantization rule $e g=2\pi$ and $N\in\mathbb{Z}$.

To find the effective potential for the radion, we promote the size of the internal sphere to a field, and rescale $\tilde{g}_{\mu\nu}$ so that Eq.~\ref{metric} becomes
\begin{equation}
\label{metricansatz}
ds^2=e^{-\psi(x)/M_4}g_{\mu\nu}dx^\mu dx^\nu + e^{\psi(x)/M_4} R^2 d\Omega_2^{\,2},
\end{equation}
where $M_4^{\,2}=4\pi R^2$.  The reason for this form is that when we integrate out the extra dimensions,
\begin{equation}
S=\int d^4x\sqrt{-g}\left(\frac12 M_4^{\,2}R^{(4)}-\frac12\partial_\mu\psi\partial^\mu\psi -V_\text{eff}(\psi)\right),
\end{equation}
we end up in Einstein frame with a canonically normalized $\psi$ field, and an effective potential
\begin{equation}
\label{effpot}
V_\text{eff}(\psi)=4\pi\left(\frac{g^2 N^2}{32 \pi^2 R^2}e^{-3\psi/M_4}-e^{-2\psi/M_4}+R^2\Lambda_6 e^{-\psi/M_4}\right).
\end{equation}
The interplay of the repulsive flux term and the attractive curvature term creates a minimum of the potential with negative $V$.  The 6D cosmological constant term can lift this vacuum all the way to positive $V$, but too much lifting removes the minimum entirely.  At the minimum, the value of $\psi$ is given by 
\begin{equation}
e^{-\psi_\text{min}/M_4}=R^2 \Lambda_6\left(\frac{N_\text{max}}{N}\right)^2  \left(1+\sqrt{1-\left(\frac{N}{N_\text{max}}\right)^2}\,\right),
\end{equation}
and the effective potential is 
\begin{equation}
\label{Vmin}
V_\text{eff}\left(\psi_\text{min}\right)=4\pi R^4\Lambda_6^{\,\,2}\left(\frac{N_\text{max}}{N}\right)^2\left[1-\frac23\left(\frac{N_\text{max}}{N}\right)^2\left(1+\left[1-\left(\frac{N}{N_\text{max}}\right)^2\right]^{3/2}\right)\right],
\end{equation}
where $N_\text{max}=4\pi\sqrt{2/3 g^2 \Lambda_6}$.  
To have a minimum at all, $N$ must be less than $N_\text{max}$. When $N = \sqrt{3/4} N_\text{max}$,  the minimum has $V_\text{eff} = 0$, which corresponds to 4D Minkowski space.  For smaller $N$, i.e.~$1\leq N<\sqrt{3/4} N_\text{max}$, the minimum has negative $V_\text{eff}$, which corresponds to 4D AdS space.  For larger $N$, i.e.~$\sqrt{3/4} N_\text{max} < N < N_\text{max}$, the minimum has positive $V_\text{eff}$, which corresponds to 4D dS space.  The Minkowski and AdS vacua can only decay by flux tunneling to smaller $N$; the dS vacuum can also decompactify.\footnote{Tunneling all the way to $N=0$ corresponds to the formation of a bubble of nothing, as in \cite{BlancoPillado:2010df, Yang:2009wz}.}  We now use our freedom to redefine $R$ by shifting $\psi$ to set
\begin{equation}
\label{Rdef}
R=\frac{1}{\sqrt{2 \Lambda_6}}.
\end{equation}
This is a convenient choice because it means $\psi_\text{min}=0$ in the Minkowski vacuum.

If we want transitions between different flux vacua, we need to remove flux, and if we want to remove flux, we need magnetically charged objects. While no such objects appear explicitly in the action, Eq.~\ref{eq:EMaction}, they are there nevertheless. Just as the Lagrangian of 4D Einstein-Maxwell theory encompasses magnetically charged black holes, so 6D Einstein-Maxwell theory encompasses magnetically charged black two-branes.  BSV \cite{BlancoPillado:2009di} argue that transitions employ the lightest possible magnetically charged object---an extremal two-brane---which has tension
\begin{equation}
\label{tension}
T_1 =  \frac{2 g}{\sqrt{3}}.
\end{equation}
This brane sits at a fixed position in the extra dimensions and forms a spherical bubble in the extended directions. It unravels one unit of flux threading the internal manifold, so that on the interior of the spherical brane there are only $N-1$ units of flux and the (four-dimensional) cosmological constant is reduced.

To calculate the tunneling rate, BSV \cite{BlancoPillado:2009di} treat the problem in the probe-brane approximation, where there is no back-reaction of the brane on the size or shape of the extra dimensions---$\psi$ is held fixed at its vacuum value.  As a consequence, the bubble wall tension, $\sigma_1$, is just the same as the brane tension, $T_1$, and, up to corrections from four-dimensional gravity, Eq.~\ref{eq:2} can be applied. 

Multijump transitions are mediated by stacks of such branes.  In a Minkowski background, extremal branes do not interact; this means that the tension of a stack of $n$ such branes scales linearly,
\begin{equation}
\label{TnT}
T_n =  nT_1.
\end{equation}
This formula remains true in a compactified space so long as the horizon size of the stack is much smaller than the size of the extra dimensions. In the probe-brane approximation, $\sigma_n = T_n$, and the bubble wall tension grows linearly with $n$. 

\subsection{Better than linear}

\begin{figure}
\begin{center}
\includegraphics[width=90mm]{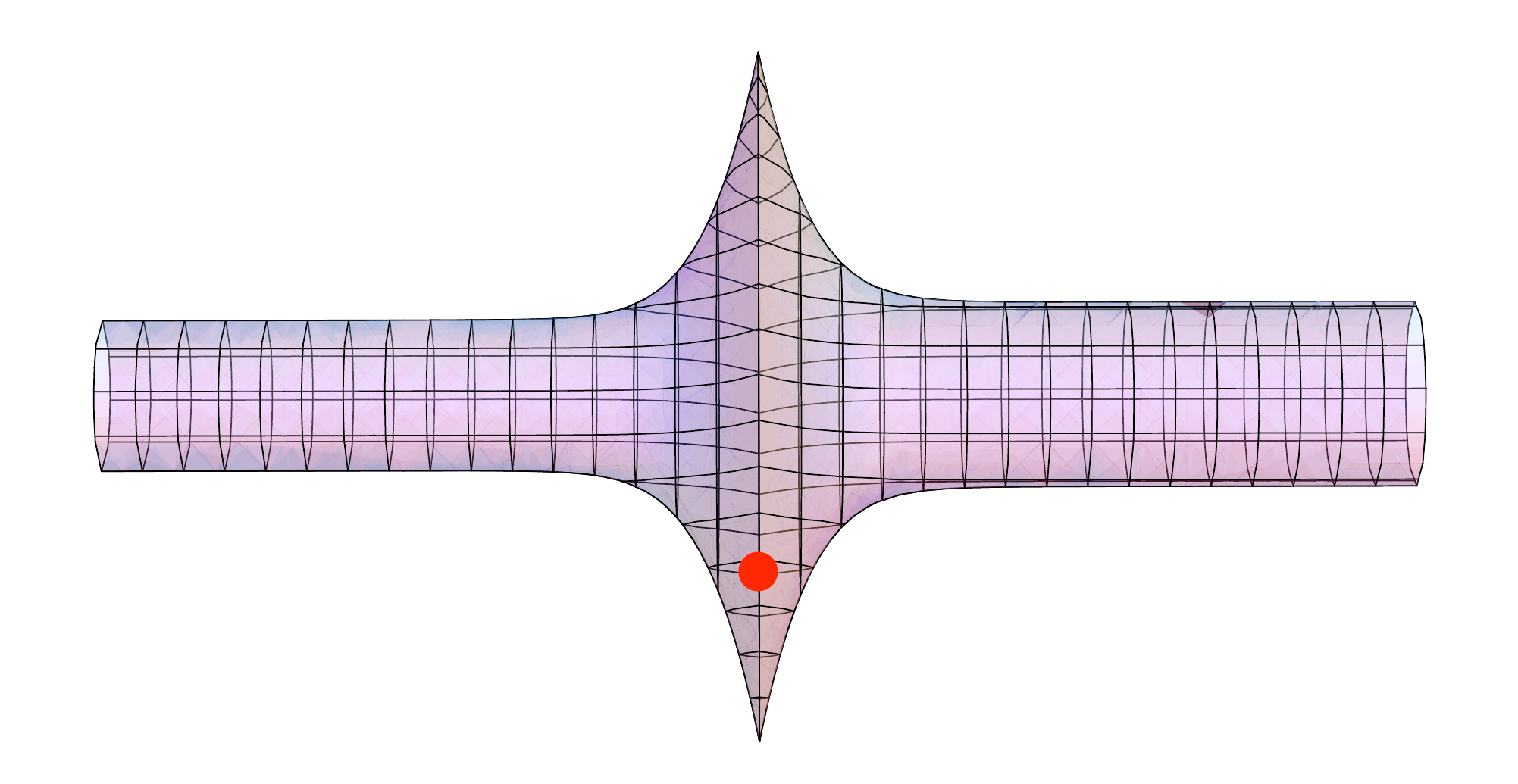}
\caption{The two-brane (indicated by a red dot) is codimension one in the extended directions, and lives at a point in the extra dimensions. It sources the radion, so that the size of the extra dimensions swells nearby.  Including this back-reaction can dramatically expedite tunneling events mediated by stacks of branes.}
\label{stretch}
\end{center}
\end{figure}

But we can do better. It was argued by Yang \cite{Yang:2009wz} that allowing the size of the extra dimensions to vary near the brane will lower the bubble wall tension. If it can be treated as infinitely thin, the two-brane contributes a term to the action proportional to its surface area as measured by the 6D theory,
\begin{equation}
S_\text{brane tension} =- T\int_\Sigma \sqrt{-\gamma} d^3\xi,
\end{equation}
where $\gamma$ is the metric induced on the brane by $G_{AB}$.  Plugging in the metric ansatz of Eq.~\ref{metricansatz} gives
\begin{equation}
\label{redshifts}
S_\text{brane tension}= -T e^{-3 \psi(\xi)/2 M_4} \sqrt{-g} d^3\xi, 
\end{equation}
where the surface area is now measured in the effective 4D theory.   As the extra dimensions swell, as in Fig.~\ref{stretch}, the effective tension of the two-brane red-shifts away. The probe-brane approximation, which amounts to freezing $\psi$ during tunneling, is like taking the straight red path instead of the  loopy blue path in Fig.~\ref{fig3}. 

We can now make the analogy with our toy model explicit.  The radion $\psi$ is the $y$ field, in the sense that both the effective potential of the radion and the effective tension of the brane asymptote to zero as $\psi\rightarrow\infty$.  Nucleating a brane is moving one rung down the ladder of minima in the $x$ direction; nucleating a stack of branes is a giant leap.  

We would like to show that giant leaps can sometimes be the dominant decay mode. To do so, we will restrict our attention to the most promising parameter regime, which in the last section we saw to be tunneling out of the $E=0$ vacuum, and we will treat it in the thin-wall approximation.

We will additionally treat the thickness of the brane as the smallest thing in the problem. This leads to a hierarchy: the radius of the bubble is much greater than the distance over which the radion varies, which in turn is much greater than the thickness of the stack of branes.

The instanton now breaks cleanly into three parts: first the $\psi$ field deforms out to some value $\psi_\text{max}$, then the flux jumps $n$ units across the brane, then the $\psi$ field comes back to its new minimum along the new potential,
\begin{equation}
\label{sigma}
\sigma_n = \int_\text{out}\!\! \sqrt{2 V_N(\psi) }\,\,d\psi + T_n e^{-3 \psi_\text{max}/2 M_4}  + \int_\text{back} \!\!\sqrt{2 V_{N-n}(\psi) }\,\,d\psi,
\end{equation}
where $V_N$ is the effective potential for $\psi$ in Eq.~\eqref{effpot} with $N$ units of flux. We take $N=\sqrt{3/4} N_\text{max}$ so that $E = V_N(\psi = 0) = 0$. 

Finally, as we are in the thin-wall limit, we will approximate the integral `back' by the integral `out'. Equation~\ref{sigma} becomes
\begin{equation}
\sigma_n = n T_1  e^{-3 \psi_\text{max}/2 M_4}  + \frac{8 \sqrt{\pi}}{3} M_4 \left(2+e^{-3 \psi_\text{max}/2 M_4}-3 e^{-\psi_\text{max}/2 M_4}\right). \label{eq31}
\end{equation}
In the probe-brane approximation, $\psi_\text{max} = 0$, this reduces to $\sigma_n = n T_1$. By contrast, the optimal $\psi_\text{max}$ is given by 
\begin{equation}
e^{\psi_\text{max}/ M_4} = 1 + \frac{3 n}{8 \sqrt{\pi}} \frac{T_1}{M_4} ;
\end{equation}
as the tension grows, it is worth stretching more. Plugging back into Eq.~\ref{eq31} gives
\begin{equation}
\sigma_n = \frac{16 \sqrt{\pi}}{3} M_4 \left[ 1 -  \left(1 + \frac{3n}{8 \sqrt{\pi} } \frac{T_1}{M_4} \right)^{-\frac{1}{2}} \, \right]. \label{sigman}
\end{equation}
In six-dimensional Minkowski spacetime, the branes are noninteracting: their magnetic repulsion cancels their gravitational attraction. But in a compact space the branes couple to the radion and, through the radion, to each other \cite{Garriga:2003gv}; this coupling is attractive. The brane binding energy means that,  just as in the toy model, the bubble wall tension grows slower than linearly with $n$. This is good news for giant leaps.

In order to find the tunneling exponent, we also need to know $\epsilon_n$, which we find by inserting $N=\sqrt{3/4}N_\text{max}-n$ into Eq.~\ref{Vmin}. More good news: $\epsilon_n$ actually grows faster than linearly with $n$.   

\begin{figure}
\begin{center}
\includegraphics[width=5in]{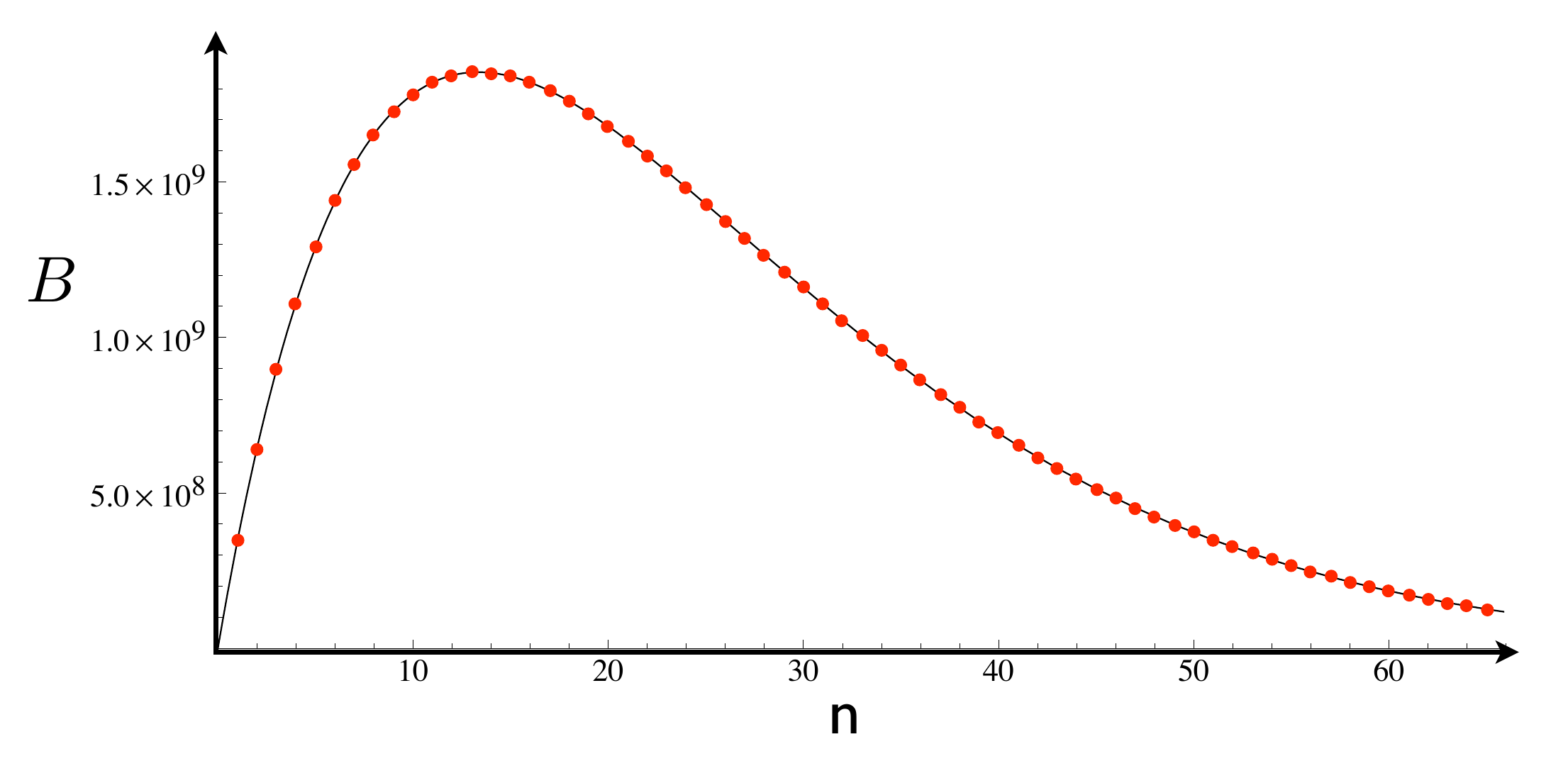}
\caption{The thin-wall approximation to the tunneling exponent $B$ to jump $n$ units away from the $E=0$ vacuum.  We have set $g^2=2 \pi^2$  and $\Lambda_6=10^{-4}$, which makes $N=200$.  It is easier to jump by fifty-two branes than jump by one. Changing either $g$ or $\Lambda_6$ changes the overall scale of $B$ (and the spacing of the dots), but not the shape of this curve.}
\label{SEofn}
\end{center}
\end{figure}

Since $B\sim\sigma_n^{\,\,4}/\epsilon_n^{\,\,3}$, these effects both contribute to giant leaps.  In fact, either one alone would suffice.  The thin-wall tunneling exponent is plotted in Fig.~\ref{SEofn}. The dots represent tunneling by integer values of the flux and, following BSV \cite{BlancoPillado:2009di}, we have chosen $g^2=2 \pi^2$ and $\Lambda_6=10^{-4}$, which puts the Minkowski vacuum at $N=200$.
In this thin-wall limit, it is easier to nucleate a stack of branes, fifty-two thick, than to nucleate just one---this vacuum prefers to tunnel a quarter of the way down the landscape rather than to its nearest neighbor.  

Let's revisit the five approximations we make to arrive at Fig.~\ref{SEofn}:

The thickness of a two-brane is $r_0=\sqrt{3}g /8\pi $ \cite{Gregory:1995qh}, so the thickness of a stack is $nr_0$.  Our first approximation is that this is much less than the bubble wall thickness, which is set by the distance scale over which the $\psi$ field varies. This allows us to use the membrane action of Eq.~\ref{redshifts}.  The bubble wall thickness can be estimated in the context of the thin-wall limit by integrating the equation of motion $\psi'=\sqrt{2V(\psi)}$ through the wall.  For our parameters, it can be checked that this is a good approximation.

Second, we approximate the thickness of the stack to be also much smaller than the size of the extra dimensions.  We need this in order to fit the branes comfortably, and to treat them as noninteracting, so that we can use Eq.~\ref{TnT}.  This can be checked to be an adequate approximation, and indeed is aided by the swelling of the extra dimensions in the vicinity of the stack.

Third, we ignore the four-dimensional curvature of the AdS true vacuum, which was shown by Coleman and de Luccia \cite{Coleman:1980aw} to impede tunneling.  Though these corrections are easy enough to include, they are negligible in our case, since the size of the bubble ($\sim 3 \sigma_n/\epsilon_n$) is much smaller than the AdS curvature length ($M_4 \sqrt{3/\epsilon_n}$).

The fourth approximation is that the wall is thin (which also justifies treating the integral `back' as being equal to the integral `out'). This approximation is our weakest. It is also uncontrolled---the toy model we have chosen, the 6D Einstein-Maxwell model, is too austere to be able to tune the wall thickness, since both $T_1$ and $\epsilon_1$ are controlled by the same parameter, $g$. The reason we nevertheless work in the thin-wall limit is that it is the clearest context in which to provide an intuitive understanding of the phenomenon of giant leaps. To investigate giant leaps away from the thin-wall limit requires a numerical analysis, which will appear in an upcoming work. 

The probe-brane approximation freezes the size and shape of the extra dimensions. In this section, we have partially moved away from this limit by unfreezing the volume modulus, but, and this is the fifth approximation, we keep the shape moduli fixed (equivalent to treating the brane as smeared round the extra dimensions). This approximation overestimates $B$. By optimizing over the size of the sphere we have greatly hastened decay to distant minima; optimizing also over the shape will only make the false vacuum decay faster and leap farther.

\section{Landscapes of Flux Vacua}  \label{secIV}

In Sec.~\ref{secII} we constructed a two-field model with a crucial ingredient, a direction in which the potential asymptotes to zero, and we showed that this theory exhibits giant leaps. Then in Sec.~\ref{secIII}, we found the same ingredient in a simple flux compactification---the effective potential for the volume modulus also asymptotes to zero. We showed that this theory, too, exhibits giant leaps in the thin-wall limit.  Dine and Seiberg argued that such a run-away direction in the effective potential, asymptoting to zero, is standard in all compactifications, arising as the coupling goes to zero or the warping goes to infinity \cite{Dine:1985he, Giddings:2003zw}. Since this crucial ingredient appears in all flux compactifications, there is good reason to believe that giant leaps are generic. 

As such, and as they are most important near $E=0$, it is natural to wonder if giant leaps have anything to say about the cosmological constant.  We find that they are relevant in two ways, both problematic: first, by opening up a new decay route, they make it harder to build long-lived vacua with small cosmological constants; and second, such vacua can be leapt over rather than landed in, complicating landscape-style resolutions of the cosmological constant problem.

\subsection{Leaping out of a Vacuum}

If you only knew about small steps, and didn't know that, sometimes, $B_n \ll B_1$, then you would underestimate decay rates, and overestimate lifetimes. Worse, you would be most wrong exactly when you are most interested in being most right: near $E=0$. Vacua that appear acceptably stable in the probe-brane approximation may actually decay unacceptably fast. Fortunately, we can give a surprisingly general characterization of the rate of decay by giant leaps. 

The probe-brane approximation, keeping $\psi$ fixed at its vacuum value, is an overestimate of $B$ that becomes increasingly accurate for the lightest branes. Let's consider the opposite limit, the heavy-brane limit. Let's have $\psi$ deform all the way out to infinity in the vicinity of the stack. This too will be an overestimate of $B$, but this time one that becomes increasingly accurate for the heaviest branes. In this approximation the tension of the stack is completely redshifted away---the only contribution to $\sigma$ is from the barrier penetration of $\psi$, which in the thin-wall limit is $\sigma = 2 \int d \psi \sqrt{2 V(\psi)}$, where $\psi$ runs between its minimum value and infinity. For the heaviest branes,  $\sigma$ approaches a constant independent of $T$, and this constant is precisely twice the tension of the wall required to decompactify. Sure enough, in the 6D Einstein-Maxwell theory, as $n \rightarrow \infty$ in Eq.~\ref{sigman}, $\sigma$ approaches $16\sqrt\pi M_4/3$, which is twice the decompactification integral. 

Armed with this upper bound, we can make a back-of-the-envelope calculation of the decay rate for any flux compactification landscape. For any transition, in any flux compactification, all we need to know is $\epsilon$ and the radion potential and we can give an upper bound on the lifetime, without knowing anything about the tension of the mediating brane. 

Let's see how this works in the KKLT compactification of type IIB string theory \cite{Kachru:2003aw}, which gives rise to an effective four-dimensional theory with a small positive cosmological constant.  Several decay modes have been considered already: the original paper considered decompactification, and then Frey, Lippert and Williams~\cite{Frey:2003dm} considered brane-flux annihilation in the manner of KPV \cite{Kachru:2002gs}.  In this type of transition, the $\overline{\text{D3}}$-branes, sitting at the tip of the throat, blow up into an NS5-brane, which removes the antibrane flux on the inside of the bubble, reducing the effective four-dimensional cosmological constant.  Frey \emph{et al.}~\cite{Frey:2003dm} treated this process in the probe-brane limit.  They found, depending on model parameters, that brane-flux annihilation can beat decompactification, and they calculated values of $B$ around $10^{18}$ to $10^{19}$.


Let's move away from the probe-brane limit.  As we've just argued, in the heavy-brane thin-wall limit, we can approximate $\sigma\sim2\sigma_\text{KKLT} \equiv 2\int\!\!\sqrt{2V_\text{KKLT}}$, where $V_\text{KKLT}$ is the potential for the volume modulus, so that $\sigma\sim8\times10^{-8}M_4^{\,3}$.  Here, $\epsilon$ is given by the scale of the uplifting done by the antibranes, so that $\epsilon\sim 2\times10^{-15}M_4^{\,4}$.  Putting these estimates into Eq.~\ref{eq:2} gives $B\sim6\times10^{17}$. Including thick-wall and gravitational corrections will raise this value, but the fact that it is competitive with the probe-brane approximation testifies to the importance of unfreezing the radion during this transition.

And what about other transitions?
In addition to KPV-style transitions, there are a large number of possible decays that take place in the body of the Calabi-Yau manifold where fluxes thread various cycles.  These transitions are also mediated by the nucleation of charged branes and again we can apply our estimate.    For all these decays, $\sigma$ is roughly $2\sigma_\text{KKLT}$. Therefore, any transition with large enough $\epsilon$ would pose a threat to the stability of the KKLT vacuum.  Since $B\sim10^3$ corresponds to lifetimes comparable to the age of the Universe, a transition with roughly $\epsilon\,\gsim\,10^{-10}M_4^{\,4}$ would invalidate the construction.

\subsection{Leaping over a Vacuum}

Anthropic resolutions of the cosmological constant problem require a vast discretuum of finely-spaced minima and a mechanism to populate the habitable ones near $E=0$.  Bousso and Polchinski constructed such a landscape in M theory \cite{Bousso:2000xa} out of flux threading hundreds of different three-cycles of an internal manifold.  Just as in 6D Einstein-Maxwell theory, transitions occur by the nucleation branes---here, M5-branes.  Whatever mechanism you use to stabilize the volume modulus, its effective potential is guaranteed to go to zero in the large volume limit \cite{Dine:1985he}.  We therefore expect giant leaps near $E=0$.

The problem is that these giant leaps can leap you straight past the habitable vacua.  The Universe starts with a large cosmological constant and then sheds flux---at first one brane at a time, but then, as it nears $E=0$, whole stacks at once.  The habitable vacua we were hoping to populate are exactly the ones that get skipped.

What this means for landscape-style resolutions of the cosmological constant problem is unclear. Though giant leaps make landing in a vacuum with a small cosmological constant less likely, they do not make it impossible.  There will be fewer bubbles of such vacua than before, and in that sense we have added to the burden of the selection criterion.  But there are still some, and a robust enough anthropic cut will still be able to pick them out. 

\section{Summary}

We have found giant leaps in a toy two-field model and, in the thin-wall approximation, in the landscape of 6D Einstein-Maxwell theory, and have argued for their importance in more general flux compactifications.  These systems share a runaway `decompactification' direction in which the potential goes asymptotically to zero, and unfreezing this `radion' mode encourages decays to wildly distant minima. This effect is strongest for vacua with a small cosmological constant and, as such, could provide the dominant decay mode of our false vacuum, so that our world will end not with a small step, but with a giant leap. 

\section*{Acknowledgements}
Thanks to Matt Johnson, Nemanja Kaloper, Igor Klebanov,  Herman Verlinde, and, particularly, Paul Steinhardt. 


\begin{thebibliography}{99}

\bibitem{Coleman:1977py}
  S.~Coleman,
  ``The Fate Of The False Vacuum. 1. Semiclassical Theory,''
  Phys.\ Rev.\  D {\bf 15}, 2929 (1977)
  [Erratum-ibid.\  D {\bf 16}, 1248 (1977)].

\bibitem{BlancoPillado:2009di}
  J.~J.~Blanco-Pillado, D.~Schwartz-Perlov and A.~Vilenkin,
  ``Quantum Tunneling in Flux Compactifications,''
  JCAP {\bf 0912}, 006 (2009)
  [arXiv:0904.3106 [hep-th]].
  
\bibitem{Freund:1980xh}
  P.~G.~O.~Freund and M.~A.~Rubin,
  ``Dynamics Of Dimensional Reduction,''
  Phys.\ Lett.\  B {\bf 97}, 233 (1980).

\bibitem{Tye:2009rb}
  S.~H.~Tye and D.~Wohns,
  ``Resonant Tunneling in Scalar Quantum Field Theory,''
  arXiv:0910.1088 [hep-th].

\bibitem{Brown:2007zzh}
  A.~R.~Brown, S.~Sarangi, B.~Shlaer and A.~Weltman,
  ``A Wrinkle in Coleman - De Luccia,''
  Phys.\ Rev.\ Lett.\  {\bf 99}, 161601 (2007)
  [arXiv:0706.0485 [hep-th]];
  A.~R.~Brown,
  ``Brane tunneling and virtual brane-antibrane pairs,''
  PoS CARGESE2007, 022 (2007)
  [arXiv:0709.3532 [hep-th]].
  
\bibitem{Feng:2000if}
  J.~L.~Feng, J.~March-Russell, S.~Sethi and F.~Wilczek,
  ``Saltatory relaxation of the cosmological constant,''
  Nucl.\ Phys.\  B {\bf 602}, 307 (2001)
  [arXiv:hep-th/0005276].
  
\bibitem{Dienes:2008qi}
  K.~R.~Dienes and B.~Thomas,
  ``Cascades and Collapses, Great Walls and Forbidden Cities: Infinite Towers
  of Metastable Vacua in Supersymmetric Field Theories,''
  Phys.\ Rev.\  D {\bf 79}, 045001 (2009)
  [arXiv:0811.3335 [hep-th]].

  
\bibitem{Aguirre:2009tp}
  A.~Aguirre, M.~C.~Johnson and M.~Larfors,
  ``Runaway dilatonic domain walls,''
  Phys.\ Rev.\  D {\bf 81}, 043527 (2010)
  [arXiv:0911.4342 [hep-th]].

\bibitem{Johnson:2008vn}
  M.~C.~Johnson and M.~Larfors,
  ``An obstacle to populating the string theory landscape,''
  Phys.\ Rev.\  D {\bf 78}, 123513 (2008)
  [arXiv:0809.2604 [hep-th]].

\bibitem{Cvetic:1994ya}
  M.~Cvetic and H.~H.~Soleng,
  ``Naked singularities in dilatonic domain wall space times,''
  Phys.\ Rev.\  D {\bf 51}, 5768 (1995)
  [arXiv:hep-th/9411170].
  

\bibitem{Douglas:2006es}
  M.~R.~Douglas and S.~Kachru,
  ``Flux compactification,''
  Rev.\ Mod.\ Phys.\  {\bf 79}, 733 (2007)
  [arXiv:hep-th/0610102].
  
\bibitem{BlancoPillado:2010df}
  J.~J.~Blanco-Pillado and B.~Shlaer,
  arXiv:1002.4408 [hep-th].

\bibitem{Yang:2009wz}
  I.~S.~M.~Yang,
  ``Stretched extra dimensions and bubbles of nothing in a toy model
  landscape,''
  arXiv:0910.1397 [hep-th].
  
\bibitem{Garriga:2003gv}
  J.~Garriga and A.~Megevand,
  ``Coincident brane nucleation and the neutralization of Lambda,''
  Phys.\ Rev.\  D {\bf 69}, 083510 (2004)
  [arXiv:hep-th/0310211].
  
\bibitem{Gregory:1995qh}
  R.~Gregory,
  Nucl.\ Phys.\  B {\bf 467}, 159 (1996)
  [arXiv:hep-th/9510202].
  
\bibitem{Coleman:1980aw}
  S.~R.~Coleman and F.~De Luccia,
  ``Gravitational Effects On And Of Vacuum Decay,''
  Phys.\ Rev.\  D {\bf 21}, 3305 (1980).
  
\bibitem{Dine:1985he}
  M.~Dine and N.~Seiberg,
  ``Is The Superstring Weakly Coupled?,''
  Phys.\ Lett.\  B {\bf 162}, 299 (1985).

\bibitem{Giddings:2003zw}
  S.~B.~Giddings,
  ``The fate of four dimensions,''
  Phys.\ Rev.\  D {\bf 68}, 026006 (2003)
  [arXiv:hep-th/0303031].
  
\bibitem{Kachru:2003aw}
  S.~Kachru, R.~Kallosh, A.~D.~Linde and S.~P.~Trivedi,
  ``De Sitter vacua in string theory,''
  Phys.\ Rev.\  D {\bf 68}, 046005 (2003)
  [arXiv:hep-th/0301240].

\bibitem{Frey:2003dm}
  A.~R.~Frey, M.~Lippert and B.~Williams,
  ``The fall of stringy de Sitter,''
  Phys.\ Rev.\  D {\bf 68}, 046008 (2003)
  [arXiv:hep-th/0305018].

\bibitem{Kachru:2002gs}
  S.~Kachru, J.~Pearson and H.~L.~Verlinde,
  ``Brane/Flux Annihilation and the String Dual of a Non-Supersymmetric Field
  Theory,''
  JHEP {\bf 0206}, 021 (2002)
  [arXiv:hep-th/0112197].
  
\bibitem{Bousso:2000xa}
  R.~Bousso and J.~Polchinski,
  ``Quantization of four-form fluxes and dynamical neutralization of the
  cosmological constant,''
  JHEP {\bf 0006}, 006 (2000)
  [arXiv:hep-th/0004134].

\end{thebibliography}
\end{document}